\documentclass[twocolumn,showpacs,showkeys,preprintnumbers,amsmath,amssymb]
              {revtex4}

\usepackage{graphicx}               
\usepackage{dcolumn}                
\usepackage{bm}                     

\begin{document}

\newcommand{\ba}{\begin{array}}
\newcommand{\ea}{\end{array}}
\newcommand{\bc}{\begin{center}}
\newcommand{\ec}{\end{center}}
\newcommand{\nen}{\nonumber}
\newcommand{\eps}{\epsilon}
\newcommand{\dbar}{{\delta \!\!\!\! \smallsetminus}}
\newcommand{\tr}{\textrm{ tr}}
\newcommand{\sumint}{{\textstyle \sum}\hspace{-2.7ex}\int}
\newcommand{\atanh}{\textrm{atanh}}

\newcommand{\UA}{\uparrow}
\newcommand{\DA}{\downarrow}

\newcommand{\x}{\mathbf{x}}
\newcommand{\ist}{\!=\!}
\newcommand{\ket}[1]{{\left| #1 \right>}}
\newcommand{\bra}[1]{{\left< #1 \right|}}
\newcommand{\bracket}[2]{{\left< {#1} \left| {#2} \right.\right>}}
\newcommand{\bigbracket}[2]{{\big< {#1} \big| {#2} \big>}}
\newcommand{\Bigbracket}[2]{{\Big< {#1} \Big| {#2} \Big>}}
\newcommand{\bigket}[1]{{\big| #1 \big>}}
\newcommand{\bigbra}[1]{{\big< #1 \big|}}
\newcommand{\Bigket}[1]{{\Big| #1 \Big>}}
\newcommand{\Bigbra}[1]{{\Big< #1 \Big|}}
\newcommand{\me}[3]{{\left< {#1} \left| {#2} \right| {#3} \right>}}
\newcommand{\bigme}[3]{{\big< {#1} \big| {#2} \big| {#3} \big>}}
\newcommand{\bigmeS}[3]{{\big< {#1} \big| {#2} \big| {#3} \big>_S}}
\newcommand{\Bigme}[3]{{\Big< {#1} \Big| {#2} \Big| {#3} \Big>}}
\newcommand{\green}[1]{{ \left<\!\left< {#1} \right>\!\right> }}
\newcommand{\biggreen}[1]{{ \big<\!\big< {#1} \big>\!\big> }}
\newcommand{\Biggreen}[1]{{ \big<\!\big< {#1} \big>\!\big> }}

\newcommand{\define}{\stackrel{\textrm{\tiny def}}{=}}
\newcommand{\mustbe}{\stackrel{!}{=}}
\newcommand{\n}{\Hat{n}}
\newcommand{\Dint}{\int\;{\cal D}[\theta,\Phi]\;}

\newcommand{\Eq}[1]{{Eq.~(\ref{#1})}}
\newcommand{\EQ}[1]{{Equation~(\ref{#1})}}
\newcommand{\av}[1]{{\left<{#1}\right>}}

\newcommand{\cdag}{c^\dagger}
\newcommand{\cnod}{c^{\phantom{\dagger}}}
\newcommand{\ctdag}{\tilde c^\dagger}
\newcommand{\ctnod}{\tilde c^{\phantom{\dagger}}}

\preprint{} 

\title{Effective spinless fermions in the strong coupling Kondo model}

\author{Winfried Koller, Alexander Pr\"ull, Hans Gerd Evertz, and Wolfgang von der Linden}
 \affiliation{Institut f\"ur Theoretische Physik,
Technische Universit\"at Graz, Petersgasse 16, A-8010 Graz, Austria.}
\email{koller@itp.tu-graz.ac.at}

\date{June 4, 2002}

\begin{abstract}
Starting from the two-orbital Kondo-lattice model with classical $t_{2g}$ spins,
an effective spinless fermion model is derived for strong Hund coupling $J_H$
with a projection technique. The model is studied by Monte Carlo simulations and
analytically using a uniform hopping approximation.
The results for the spinless fermion model
are in remarkable agreement with those of the original Kondo-lattice model,
independent of the carrier concentration, and even for moderate Hund coupling $J_H$.
Phase separation, the phase diagram in uniform hopping approximation, as well as
spectral properties including the formation of a pseudo-gap are
discussed for both the Kondo-lattice and the  effective spinless fermion model
in one and three dimensions.
\end{abstract}



\maketitle

\section{Introduction}                                  \label{sec:intro}

The study of manganese oxides such as La$_{1-x}$Sr$_x$MnO$_3$ (LSMO)
and La$_{1-x}$Ca$_x$MnO$_3$ (LCMO) has attracted considerable attention since the discovery of
colossal magnetoresistance in these compounds
\cite{dagotto01:review,oles00:_magnet_orbit_order_cuprat_mangan}. These materials crystallize in
the perovskite-type lattice structure where the crystal field partially lifts the degeneracy of the
manganese d-states. The energetically favorable three-fold degenerate $t_{2g}$ levels are populated
with  localized electrons, which according to Hund's rule form localized $S=3/2$ spins. The
electronic configuration of the Mn$^{3+}$ ions is $t_{2g}^3e_g^1$ with one electron in
the $e_g$ orbital, which is missing in the Mn$^{4+}$ ions. The $e_g$ electrons can move between
neighboring Mn ions mediated by bridging $O^{2-}$ $2p$ orbitals. The interplay of electronic, spin,
and orbital degrees of freedom along with the mutual interactions, such as the strong Hund
coupling $J_H$ of the itinerant electron to localized $t_{2g}$ spins, Coulomb correlations, and
electron-phonon coupling leads to a rich phase diagram including antiferromagnetic insulating,
ferromagnetic metallic, and charge ordered domains \cite{proceedings98}.
Charge carriers moving in the spin and orbital background
show interesting dynamical features \cite{horsch99,bala02}.
The electronic degrees of
freedom are generally treated by a Kondo-lattice model, which in the strong Hund coupling limit
is commonly referred to as double-exchange (DE) model, a term first coined by Zener\cite{zener51}.

Monte Carlo (MC) simulations have contributed significantly to our understanding of the manganites.
Intense MC simulations for the DE model have been performed by Dagotto {\em et
al} \cite{dagotto98:_ferrom_kondo_model_mangan} and Furukawa \cite{furukawa98} in the space of the
classically treated $t_{2g}$ spins. Static and dynamical observables of the Kondo model have been
determined \cite{yunoki98:_static_dynam_proper_ferrom_kondo}. These MC simulations gave
first theoretical indications of phase separation (PS)~\cite{yunoki98:_phase} in manganite models.
Preliminary studies have been performed to analyze  the importance of nearest neighbor Coulomb
repulsion in the two-orbital DE model~\cite{hotta00:coo_ps_nn_coulomb} as well as  the importance
of classical phonons~\cite{yunoki98:_phase_separ_induc_orbit_degrees}.

Many publications are based on the $J_H=\infty$ limit.
Here we propose an effective spinless fermion model for the strong coupling
limit of the Kondo-lattice Hamiltonian,
which is equally simple as the $J_H=\infty$ limit but which still contains the
crucial physical ingredients of finite $J_H$.
The dynamic variables are $e_g$ electrons with spins parallel to the $t_{2g}$ spins
at the respective sites.
The influence of antiparallel spins is accounted for by the effective Hamiltonian.
The derivation of the model is based on a projection technique, analogous to
the derivation of the $tJ$ model from the Hubbard model.
The role of the Hubbard~$U$ is played by $J_H$ which couples to the classical
$t_{2g}$ spins.
In contrast to the $tJ$ model, the high-energy subspace is thus controlled
by classical variables and consequently the resulting model is much simpler
than the $tJ$ model.
For a given $t_{2g}$ spin configuration, the resulting Hamiltonian is a
{\em one-particle} operator.
Its electronic trace can be evaluated analytically, once the one-particle
energies are known, leading to an effective action for the $t_{2g}$ spins,
which can be simulated by Monte Carlo techniques.

The obvious advantage of this approach is the reduction of the
dimension of the Hilbert space. This can be exploited in MC simulations by
going to larger systems and/or additional degrees of freedom.

For a large range of parameters, the effective spinless
fermion model is found to yield very satisfactory
results and to perform much better than the rough $J_H \to \infty$
approximation.
We compare spin- and charge-correlations as well as quasi-particle
spectra of the projection approach with the full
two-spin model and with the $J_H\to\infty$ limit.

The effective model is treated without approximations by Monte Carlo
simulations as well as by a uniform hopping approximation (UHA)
capturing the essential influence of the $t_{2g}$ spins on the $e_g$
electrons.
The UHA computation can be performed  analytically, particularly in the thermodynamic limit.
Most of the UHA results are found to be in striking agreement with MC results.
We find two phase transitions as a function of the chemical potential, one close
to the empty band and the second close to a completely filled band. At each phase transition
we observe PS,
as reported for the upper transition in \cite{yunoki98:_phase}.
For a 1D-chain we derive an analytical expression for the two critical chemical potentials
at which (PS) occurs.

For the 3D Kondo-lattice model
canonical UHA results yield a phase diagram which displays various types of
antiferromagnetic (AF) order including spin-canting, as well as ferromagnetism (FM).
Our finite $J_H$ results for 3D are in close agreement with those derived in the limit of
infinite Hund coupling \cite{brink99:_DE_two_orbital}.
In the grand canonical ensemble we find, however, that only the 3D antiferromagnetic
and the 3D ferromagnetic
order prevail. The transition is again accompanied by PS.

The paper is organized as follows. In Sec.~\ref{sec:model} the
Kondo-lattice model is introduced.
By applying a projection technique in Sec.~\ref{sec:effective}
this model is mapped onto the effective spinless fermion model.
In Sec.~\ref{sec:analytic} we present the  phase diagrams and phase separation boundaries in
one and three dimensions within a uniform hopping approximation.
Results of Monte Carlo simulations for the original and the projected model are discussed in
Sec.~\ref{sec:results}. Finally, in Sec.~\ref{sec:conclusions} we summarize the key conclusions.

\section{Model Hamiltonian}                        \label{sec:model}

In this paper, we will concentrate on purely electronic ($t_{2g}, e_g$) properties,
leaving phonon degrees of freedom for further studies. As proposed by Dagotto {\it et al.}
\cite{dagotto98:_ferrom_kondo_model_mangan} and Furukawa \cite{furukawa98}, the $t_{2g}$ spins~$\mathbf
S_i$ are treated classically, which is equivalent to the limit $S\to \infty$. The spin degrees of
freedom are therefore replaced by unit vectors $\mathbf S_i$, parameterized by polar and azimuthal
angles  $\theta_i$ and $\phi_i$, respectively, that represent the direction of the $t_{2g}$ spin at
lattice site~$\x_i$. The magnitude of the spin is absorbed into the exchange couplings. It is
expedient to use the individual $t_{2g}$ spin directions as local quantization axes for the spin of the
itinerant $e_g$ electrons at the respective sites. This representation is particularly useful for
the $J_H\to\infty$ limit, but also for the projection technique, which takes spin-flip
processes for finite Hund coupling into account.

It is commonly believed that the electronic degrees of freedom are well described by
a multi-orbital Kondo-lattice model
\begin{widetext}
\begin{equation}                                           \label{eq:H}
  \hat H = -\sum_{i,j,\alpha,\beta, \sigma,\sigma'}\;
  t^{\sigma,\sigma'}_{i\alpha,j\beta}\;
  \cdag_{i\alpha\sigma}\,\cnod_{j\beta\sigma'}
  + 2 J_H \sum_{i\alpha} \n_{i\alpha\DA} %
  + J'\sum_{<ij>} \mathbf S_i \cdot \mathbf S_j\;.
\end{equation}
\end{widetext}
It consists of a kinetic term with modified transfer integrals
$t_{i\alpha,j\beta}^{\sigma,\sigma'}$, where $i(j)$ are site-, $\alpha(\beta)$ orbital-, and
$\sigma(\sigma')$ spin indices. The number of lattice sites will be denoted by $L$ and the number of
orbitals per site by $M$. The operators $\cdag_{i\alpha\sigma} (\cnod_{i\alpha\sigma})$ create
(annihilate) $e_g$-electrons at  site $x_i$ in the orbital $\alpha$ with spin parallel ($\sigma =
\UA$) or anti-parallel ($\sigma = \DA$) to the local $t_{2g}$ spin orientation~$\mathbf S_i$. The next
term describes the Hund coupling with exchange integral $J_H$. As usual,
$\n_{i\alpha\sigma}$ is the spin-resolved occupation number operator.
Usually, the Kondo-lattice Hamiltonian contains an additional term
proportional to the electron number $\hat{N}_e$,
$\hat{H}_c= - J_H \hat{N}_e$, which has been omitted in \Eq{eq:H}, as it merely results in a trivial
shift of the chemical potential $\mu\to\mu-J_H$.

The modified hopping integrals $t^{\sigma,\sigma'}_{i\alpha,j\beta}$ depend upon the geometry of the
$e_g$-orbitals and the relative orientation of the $t_{2g}$ spins:
\[
  t^{\sigma,\sigma'}_{i\alpha,j\beta} \;=\;
  t_{i\alpha,j\beta}\; u_{ij}^{\sigma,\sigma'}\;.
\]
The first factor on the RHS is given by the hopping amplitudes $t_{i\alpha,j\beta}$ which read
\begin{equation}\label{eq:hopping}
  t_{i,i+\hat z} \;=\; t \;
    \left(\ba{cc}
    0 & 0 \\ 0 & 1
    \ea \right),\;
  t_{i,i+\hat x/\hat y} =  t \;
    \left(\ba{cc}
    \tfrac{3}{4} &\mp\tfrac{\sqrt{3}}{4} \\ \mp\tfrac{\sqrt{3}}{4} & \tfrac{1}{4}
    \ea \right)
\end{equation}
as matrices in the orbital indices $\alpha,\beta=1 (2)$, corresponding to the $x^2-y^2$
($3z^2-r^2$) orbitals (see e.g. \cite{dagotto01:review}). The overall hopping strength is $t$, which
will be used as unit of energy, by setting $t=1$. The relative orientation of the $t_{2g}$ spins at site
$i$ and $j$ enters via
\begin{equation}
 \begin{aligned}
    u^{\sigma,\sigma}_{i,j} &= c_i c_j + s_i s_j \;
    e^{i\sigma (\phi_j-\phi_i)}\\
    u^{\sigma,-\sigma}_{i,j} &= \sigma(c_i s_j\;e^{-i \sigma \phi_j} -
    c_j s_i \; e^{-i\sigma \phi_i})
  \end{aligned}\;,
\end{equation}
with the abbreviations $c_j = \cos(\theta_j/2)$ and $s_j = \sin(\theta_j/2)$ and
the restriction $0\le\theta_j\le\pi$. The modified hopping
part of the Hamiltonian is still hermitian, since $u_{i,j}^{\sigma,\sigma'} = \big(u_{j,i}^{\sigma',
\sigma}\big)^*$.

Finally, \Eq{eq:H} contains a super-exchange term. The value of the exchange coupling is
$J'\approx 0.02$ \cite{dagotto01:review},
accounting for the weak antiferromagnetic coupling of the $t_{2g}$ electrons.
Here we will approximate the local $t_{2g}$ spins classically.
For strong Hund coupling $J_H\gg 1$ the electronic density of states (dos)
consists essentially of two sub-bands, a lower and an upper 'Kondo-band', split by
approximately $2J_H$. In the lower band the itinerant $e_g$ electrons move such
that their spins are predominantly parallel to the $t_{2g}$ spins, while the opposite is
true for the upper band\cite{wvdl82}. Throughout this paper, the electronic density $n$
(number of electrons per site) will be restricted to
$0\le n\le1$, i.e. predominantly the lower Kondo-band is involved.

\section{Projection Technique}                     \label{sec:effective}

The separation of energy scales~\cite{Auerbach:book} is 
a well known strategy to simplify quantum-mechanical many-body systems.
In the case of manganites, the Hund coupling $J_H$ is
known to be much greater than the other parameters $t$ and $J'$.
Consequently, the hopping to antiparallel $e_g-t_{2g}$ configurations
can be treated in second order perturbation
theory~\cite{SQShen00:pp_mf, yarlagadda01:mf_COSO}
by a projection approach.
On the low energy scale, the dynamical variables are $e_g$ electrons with spin
parallel to the local $t_{2g}$ spins.
The virtual  excitations  $(i\alpha\UA) \to (j\beta\DA) \to (i'\alpha'\UA)$,
which are mediated by the hopping matrix,
lead to an  effective spinless fermion Hamiltonian
\begin{widetext}
\begin{equation}                                       \label{eq:Hp}
  \hat H_p = -\sum_{i,j,\alpha,\beta}
    t^{\UA\UA}_{i\alpha,j\beta}\,
    \cdag_{i\alpha}\,\cnod_{j\beta}
    - \sum_{i,\alpha,\alpha'}
    \bigg(\sum_{j,\beta}
    \frac{t^{\UA\DA}_{i\alpha',j\beta}\,t^{\DA\UA}_{j\beta,i\alpha}}
    {2J_H}\bigg) \cdag_{i\alpha'}\cnod_{i\alpha}
    - \sum_{[i\ne i'],\alpha,\alpha'}\!\!\!
    \bigg(\sum_{j,\beta}
    \frac{t^{\UA\DA}_{i'\alpha',j\beta}\,t^{\DA\UA}_{j\beta,i\alpha}}
    {2J_H}\bigg) \cdag_{i'\alpha'}\cnod_{i\alpha}
    + J'\sum_{<ij>} \mathbf S_i \cdot \mathbf S_j \;.
\end{equation}
\end{widetext}
The effective Hamiltonian contains the kinetic energy of $e_g$ electrons
with spin parallel to the $t_{2g}$ spins (first term).
The kinetic energy is optimized by aligning all $t_{2g}$ spins which is the
usual ferromagnetic double exchange effect.
The second term describes an additional hybridization and favors
antiferromagnetic $t_{2g}$ spins 
leading to an effective antiferromagnetic interaction $J_{\text{eff}}$
which is generally stronger than $J'$.
The ``three-site'' hopping processes  of the third term are of minor influence.
We will see that this term is in general negligible.
On the other hand, its inclusion does not really increase the numerical effort.
\Eq{eq:Hp} is valid for arbitrary hopping matrices $t_{i\alpha,j\beta}$.
In subsequent sections, however, the
discussion will be restricted to nearest-neighbor hopping only.

The Hamiltonian \Eq{eq:Hp} constitutes a spinless fermion model,
similar to the one obtained in the $J_H\to\infty$-limit, which can
be treated numerically along the lines proposed by Dagotto and
coworkers~\cite{dagotto98:_ferrom_kondo_model_mangan} and Furukawa
\cite{furukawa98}. Finite $J_H$ values can thus be treated with the same
numerical effort as the case $J_H = \infty$.
In the MC simulations the weight for a $t_{2g}$ spin
configuration is determined by the grand canonical trace over the fermionic degrees of freedom
in the one-electron potential created by the $t_{2g}$ spins.

The obvious advantage of \Eq{eq:Hp} as compared to \Eq{eq:H} is the reduced Hilbert space.

\section{Uniform Hopping approximation}      \label{sec:analytic}
Before discussing approximation-free MC results for the effective spinless fermion model, we will
investigate the main features of the Hamiltonian~(\ref{eq:Hp}) by a uniform hopping approximation
proposed by van-den-Brink and Khomskii\cite{brink99:_DE_two_orbital}. To this end, we introduce two
different mean angles between neighboring $t_{2g}$ spins, one in $z$-direction ($\theta_z$) and one in
the $xy$-plane ($\theta_{xy}$). It should be stressed that $\theta_z$ and
$\theta_{xy}$ are relative angles between adjacent spins, with values between
$0$ and $\pi$, and are not to be confused with the polar angles $\theta_i$.
We assume that these angles are the same between all neighbor spins,
{\it i.e.} $\mathbf S_i\cdot \mathbf S_{i\pm\hat z}=\cos \theta_z$ for all lattice sites $i$ and
$\mathbf S_i\cdot \mathbf S_{i\pm\hat x}=
 \mathbf S_i\cdot \mathbf S_{i\pm\hat y}=\cos \theta_{xy}$.
The allowed spin configurations include, among others, ferro- and  antiferromagnetism as well as
spin canted states. The impact of the $t_{2g}$ spins on the hopping amplitudes simplifies to
\[
  u^{\sigma,\sigma}_z  = \cos(\frac{\theta_z}{2})\quad,\qquad
  u^{\sigma,-\sigma}_z = \sin(\frac{\theta_z}{2})\;.
\]
for hopping processes along the $z$-direction and similarly for $u^{\sigma,\sigma'}_{xy}$ for
electron motion in the $xy$-plane. The hopping matrix is now translationally invariant. The inner
product of the $t_{2g}$ spins entering the super-exchange reads
\[
\mathbf S_i\cdot \mathbf S_{i+\hat z} = \cos\theta_z = 2u_z^2-1
\]
and similarly for neighboring pairs in the $xy$-plane.
%
\subsection{Phase Separation in 1D systems} \label{sec:analytic_1D}
\begin{figure}[!ht]
\includegraphics[width=0.3\textwidth]{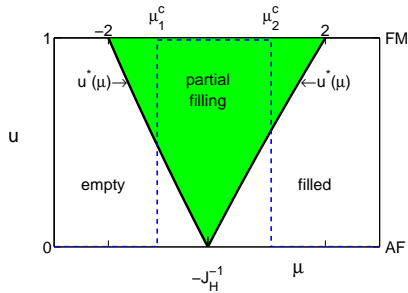}
\caption{Filling of the tight-binding band depending upon chemical potential
  and hopping amplitude. Condition for partial filling: $u>u^*(\mu)$.
The minimum free energy solution in UHA (dashed line)
exhibits a jump (phase separation)
from AF to FM order at $\mu^c_1$ and back to AF order at $\mu^c_2$. } \label{filling}
\end{figure}

\begin{figure}
\includegraphics[width=0.40\textwidth]{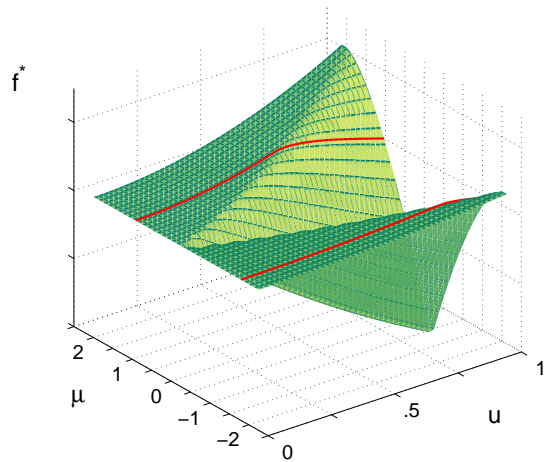}
\caption{Free energy  as a function of chemical potential and
  hopping amplitude $u$ for $T=0$, $J_H=6$ and $J'=0.02$.
  The solid lines are for $\mu^c_1$ and $\mu^c_2$, respectively.
  For the sake of clarity, $f(u=0,\mu)$ has been subtracted.}
\label{fig:free_energy.eps}
\end{figure}

%
First we consider the simplest case, namely a 1D chain in $z$-direction and we ignore the additional
three-site hopping (third term in \Eq{eq:Hp}).
At the end of this section we will show that it can indeed be neglected.
Due to the symmetry of the hopping elements the
$x^2-y^2$-orbitals form an irrelevant dispersionless band,
which will be ignored in the sequel.
The influence of the average spin orientation is captured in the uniform hopping amplitude~$u$.
Assuming periodic boundary conditions, the Hamiltonian simplifies to
\begin{equation}\label{eq:H_1D}
      \hat H = -u \sum_{\langle i j\rangle} \cdag_{i}\cnod_{j}
  - \frac{1-u^2}{J_H}\,\sum_i \cdag_{i}\cnod_{i}
  +J'L\big(2u^2-1\big)\;,
\end{equation}
where we have dropped orbital indices.
The virtual hopping processes couple merely to the density
and the dispersion of the spinless fermions is given by the shifted
tight-binding band structure 
\begin{equation}\label{eq:dispersion}
    \eps_k = -2u\cos(k) - (1-u^2)/J_H\;.
\end{equation}
The band width is $4 u$. It vanishes accordingly for AF order and reaches a maximum for FM order. In
Fig.~\ref{filling} the resulting band-filling is schematically depicted for zero temperature as
function of chemical potential and hopping amplitude.
The condition for an empty/ filled band depends on the 'effective chemical potential'
\begin{equation}\label{eq:eff_cp}
   \tilde{\mu}:=\big(\mu + \tfrac{1}{J_H}  (1-u^2)\big)/u \;.
\end{equation}
According to  $\mu< \min_k(\eps_k)$, the band is empty if
 \hbox{$\tilde{\mu}<-2$}.
For the completely filled band the condition reads \hbox{$\tilde{\mu}> 2$}.
Partial filling is
possible for intermediate values of the chemical potential ($-2< \mu <2$) if the hopping
amplitude exceeds a threshold $u^*(\mu)$.
The logarithm of the grand canonical partition function reads
\[
  \ln Y = \sum_k
  \ln\big(1+e^{-\beta(\eps_k-\mu)}\big)
  - \beta J'L\big(2u^2-1\big)\;.
\]
In the thermodynamic limit ($L\to\infty$) and for $T=0$ the free energy per lattice site is
\begin{equation}                                \label{eq:lnY}
  f=
  u E(\tilde{\mu})- u \;\tilde{\mu}\; N(\tilde{\mu}) + J'(2 u^2 - 1)\;,
\end{equation}
with $E(x)$ being the mean kinetic energy and $N(x)$ the mean particle number of a tight-binding
band with dispersion $-2\cos(k)$. For $|x|\le 2$ these quantities are
\[
  E(x) = -\frac{\sqrt{4-x^2}}{\pi}\;,\quad
  N(x) = \frac{1}{2} + \frac{\arcsin(x/2)}{\pi}\;.
\]
The kinetic energy $E(x)$ is zero for the empty band $(x<-2)$, as well as for the completely filled
band ($x>2$). The mean particle number $N(x)$ is zero if the band is empty and unity if it is
full. Fig.\ \ref{fig:free_energy.eps} shows the free energy as a function of chemical
potential and hopping amplitude~$u$. We find local minima at $u=0$ (AF order) and $u=1$ (FM
order). The kinetic terms decrease with increasing $u$, favoring FM order, while the $t_{2g}$ spin
energies increase with increasing $u$, favoring AF order. The global minimum switches from AF to
FM at the critical chemical potential $\mu=\mu^c_1$ and  back to AF at $\mu=\mu^c_2$.
The values for the critical chemical potential follow from the condition $f|_{u=0}=f|_{u=1}$,
yielding
\begin{equation}                           \label{eq:PS}
  (\mu^c + 1/J_H)\, N_0(\mu^c) + 2J' = \mu^c\,N(\mu^c)-E(\mu^c)\;,
\end{equation}
where $N_0(\mu^c)$ denotes the mean particle number for $u=0$, i.e.\ for perfect AF order. In this
case, the tight-binding band is dispersionless and  $N_0(\mu^c)$ is  either zero or one, depending
upon the actual value of the chemical potential.

For the standard parameter set $J_H=6$ and $J'=0.02$ the numerical values for $\mu^c$ are
$\mu^c_1=-1.6730$ and $\mu^c_2=1.03431$.

\begin{figure}[!ht]
\includegraphics[width=0.35\textwidth]{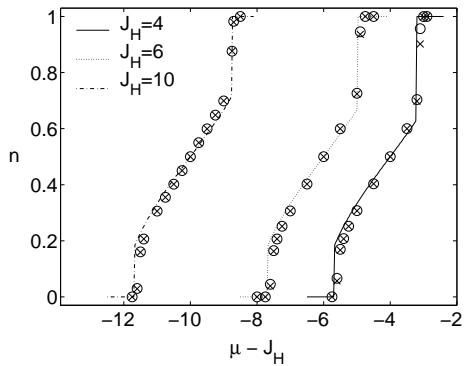}
\caption{Electron density versus chemical potential for $J_H=4, 6$, and $10$ (right to left),
 and $J'=0.02$.
 The lines correspond to UHA.
 MC results at  $\beta=50$, $L=20$ for the spinless fermion model $H_p$ (circles)
 are compared with those for the
 DE model $H$ (crosses). Error bars of the MC data are smaller than the symbols.
}
\label{mu_n_L20Jx.eps}
\end{figure}

The UHA solution corresponds to the global minimum of the free energy. Its location is
depicted in Fig.\ \ref{filling} and the corresponding densities are shown
as lines in Fig.\ \ref{mu_n_L20Jx.eps}.
For large negative chemical
potential the system is antiferromagnetic and the $e_g$ band is empty. At $\mu^c_1$, AFM domains with
zero electron density coexist with FM domains with finite density $n_1$. Increasing $\mu$  leads to
ferromagnetism, and the filling increases gradually with $\mu$,
following the tight binding formula \hbox{$n=\arccos(-\mu/2)/\pi$}.

At $\mu^c_2$, FM domains with density
$n_2>n_1$ coexist with AFM domains of density one. Finally, for $\mu>\mu^c_2$, the system jumps
back to antiferromagnetism, now at density one.
We thus see that the system exhibits phase separation.
It should be pointed out that PS is
suppressed if nearest-neighbor Coulomb repulsion among $e_g$ electrons is included into the model
\cite{kol:pru:wvl:2}.

\begin{figure}
  \includegraphics[width=0.30\textwidth]{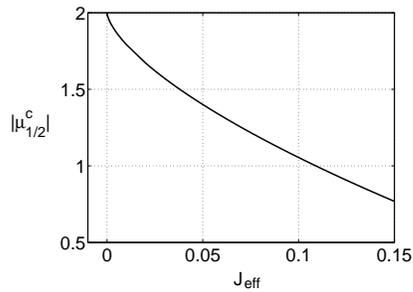}
  \caption{Dependence of the critical chemical potential $|\mu^c_{1,2}|$ on
    the effective exchange coupling $J_{\text{eff}}$.}
  \label{fig:muc}
\end{figure}

Let us now discuss the values of $\mu^c_{1,2}$ and the size of the
discontinuity in $n$.
According to \Eq{eq:PS}, the first critical chemical potential
$\mu^c_1=\mu^c(J')$, corresponding to $N_0(\mu)=0$, is independent of
the actual value of $J_H$.
Here the effective antiferromagnetic interaction $J_{\text{eff}}$ purely
stems from the superexchange coupling of the $t_{2g}$ spins, $J_{\text{eff}} = J'$.
This does, however, not mean that the Hund coupling is irrelevant for this
phase transition. On the contrary, the phase transition 
is driven by the FM tendency introduced by the Hund coupling.
The independence of $J_H$ in our results means that there are no second order
corrections to the $J_H=\infty$ limit.
The dependence of $\mu^c$ on $J'$ is depicted in Fig.\ \ref{fig:muc}.

Now we turn to the second phase transition corresponding to $N_0(\mu^c)=1$.
This transition is controlled by the stronger effective exchange coupling
\begin{equation} \label{Jeff}
  J_{\text{eff}} = J' + \tfrac{1}{2 J_H} \;,
\end{equation}
as can be seen from \Eq{eq:PS}, or directly from \Eq{eq:H_1D} at $N=1$.
Due to the particle-hole relation $N(-\mu) + N(\mu) = 1$ the
second critical chemical potential is given by $\mu^c_2 = -\mu^c(J_{\text{eff}})$,
depicted in  Fig.\ \ref{fig:muc}.
In the limit
$J_H\to\infty$ we have $\mu^c_1 = -\mu^c_2$.
The density discontinuity is $\Delta n := N(\mu^c_1)=
\Delta n(J')$ at $\mu^c_1$ and  $\Delta n := 1-N(\mu^c_2)= \Delta n(J_{\text{eff}})$ at $\mu^c_2$.
\begin{figure}
  \includegraphics[width=0.30\textwidth]{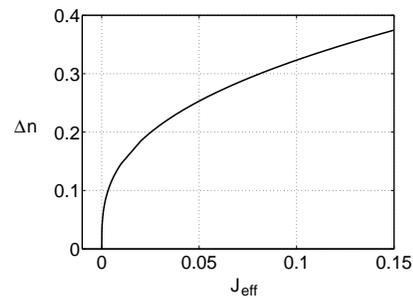}
  \caption{Discontinuity in $n$ as a function of the effective exchange
    coupling $J_{\text{eff}}$.}
  \label{n_vs_J1.eps}
\end{figure}
It is depicted in Fig.~\ref{n_vs_J1.eps}. Series expansion of \Eq{eq:PS} with respect to $\mu^c$
about $\mu^c=-2$ yields \hbox{$\Delta n=(3 J'/\pi^2)^{1/3}$}. At
$J_{\text{eff}}=0$ the slope of the curve
diverges, implying that already an infinitesimal $J_{\text{eff}}$ leads to
phase separation in this regime.

For realistic parameter values ($J_H=6$ and
$J'=0.02$) we find $\Delta n_1\approx 0.18$ and $\Delta n_2\approx 0.3$, respectively.
The second value is mainly
driven by the virtual hopping, which increases the tendency towards PS.
\vspace{5mm}

\subsection{Impact of ``three-site-terms''} \label{sec:3-site-terms}

Here we consider the impact of the additional  hopping in \Eq{eq:Hp}, which in the 1D case with one
orbital results in a next-nearest neighbor hopping
\[
  \hat H^* = -\frac{1-u^2}{2J_H}\,\sum_i\cdag_{i+2}\,\cnod_i + h.c.\qquad.
\]
Combined with the  terms of \Eq{eq:H_1D}, the resulting single-particle dispersion reads
\[
  \eps_k = -2u\cos k \;-\; \frac{1-u^2}{J_H}  \;-\;\frac{1-u^2}{J_H}\,\cos(2k)\;.
\]
In the limit $u\to 1$ we recover the original tight-binding band. The density of states
has additional van Hove
singularities. Contrary to the dispersion of \Eq{eq:H_1D}, the band width  remains
finite in the limit $u\to 0$, due to $\hat{H}^*$.

Next we derive the conditions for phase separation. In the limit $T\to0$, the free energy is given
by
\[
  f=\int_{-\infty}^\mu d\eps\;\rho_u(\eps)(\eps-\mu) \;-\;J'(2u^2-1)\;,
\]
where $\rho_u(\eps)$ denotes the density of states corresponding to $\eps_k$.
Numerical evaluation shows that the
minima of $f$ are still at $u=0$ and $u=1$. The condition for phase separation is therefore still
$f|_{u=0}=f|_{u=1}$. In principle, due to the finite width of the band at $u=0$ intermediate particle
numbers $N_0(\mu^c)$ are possible. A detailed
calculation shows, however, that for realistic parameters, only
$N_0(\mu^c)=0$ or $N_0(\mu^c)=1$ can meet the phase separation criterion.
For $N_0(\mu^c)=0$ and $N_0(\mu^c)=1$,
however, no hopping is possible and consequently the additional hopping term vanishes. Therefore,
the criterion for PS is the same as in \Eq{eq:PS} and the three-site hopping has no influence on
the critical chemical potentials~$\mu_c$ at which phase separation occurs.

In general, the modification of the bandwidth due to next-nearest neighbor hopping is small.
It has almost negligible impact on the results. On the other
hand it poses little extra-effort to include it in MC simulation.

\subsection{Phase Diagram in 3D systems}

We now turn to the 3D case. In uniform hopping approximation, the super-exchange reads
\[
\begin{aligned}
  \hat H_{se} &= J'L^3\big(2(2u_{xy}^2-1)+(2u_z^2-1)\big)\;
\end{aligned}
\]
where $L$ denotes the linear dimensions of the system. Upon substituting uniform hopping
amplitudes into \Eq{eq:Hp}, the fermionic part of the Hamiltonian can easily be diagonalized. The
one-particle energies are given by the eigenvalues of a $2\times 2$ matrix with matrix elements
\begin{equation}\label{eps}
  \begin{aligned}
    \eps_{11}(k) =&
    \frac{3}{2}\Big(-u_{xy}(\cos k_x + \cos k_y)-(1-u_{xy}^2))/J_H\Big) \\
    \eps_{12}(k) =&
    \frac{\sqrt{3}}{2}\, u_{xy}\big(\cos k_x - \cos k_y\big) =  \eps_{21}(k)\\
    \eps_{22}(k) =&
    -2u_z \cos k_z - u_{xy}\big(\cos k_x + \cos k_y\big)/2\\
    &-\big(1-u_z^2 + (1-u_{xy}^2)/2\big)/J_H
  \end{aligned}
\end{equation}
where the subscript $1$ ($2$) refers to $x^2-y^2$ ($3z^2-r^2$) orbitals. The symmetry of the
$e_g$-wavefunction has been exploited in the above expressions. As a consequence of the UHA
with two different angles $\theta_z$ and $\theta_{xy}$, virtual
hopping processes cannot induce transitions between different orbitals, and $J_H$ appears only in
the diagonal elements of the matrix.

\begin{figure}
  \includegraphics[width=0.40\textwidth]{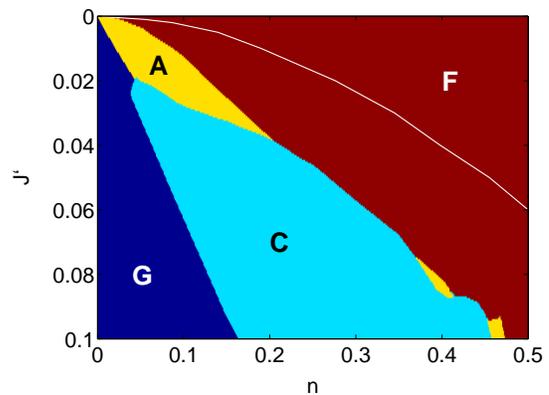}
  \caption{Canonical $T=0$ UHA phase diagram of the spinless two-orbital Kondo-lattice model
    with classical $t_{2g}$ spins for $J_H=8$.
    Depending upon the electron concentration of the canonical ensemble and the super-exchange,
    G-type, A-type, C-type antiferromagnetic phases or the
    ferromagnetic phase (F) are observed.
    Meaning of phases: G=(AF,AF); A=(FM,AF); C=(AF,FM); F=(FM,FM), where the
    first entry denotes the order in the xy-plane and the second in z-direction.
    The solid line represents the phase-boundary between G- and F-order, respectively,
    obtained in a grand canonical ensemble.
    }
  \label{phases3d_J8.eps}
\end{figure}

\begin{figure}[!ht]
  \includegraphics[width=0.35\textwidth]{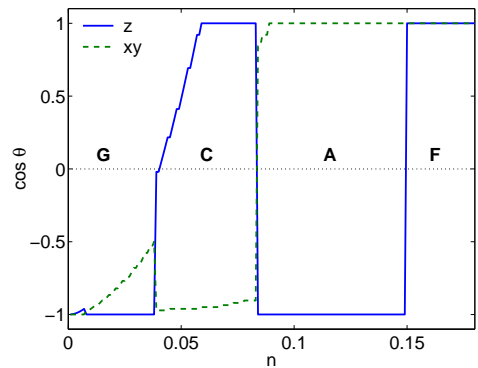}
  \caption{Evolution of the optimal angles $\theta_z$ and $\theta_{xy}$ as
    function of the electron density $n$ at $J'=0.025$, $J_H=8$.
    Further details like in Fig.\ \ref{phases3d_J8.eps}.}
  \label{tztxy_vs_y_J8S025.eps}
\end{figure}

We determine the phase diagram in the {\it canonical ensemble} at $T=0$
with respect to electron density $n$ and exchange coupling $J'$ for
fixed Hund coupling $J_H=8$,
with \Eq{eps} evaluated on a $20^3$ momentum lattice.
For each parameter set, the free energy is minimized with
respect to the hopping amplitudes. The resulting phase diagram is depicted in
Fig.~\ref{phases3d_J8.eps}. At very low doping, 3D antiferromagnetic (G) order  dominates, irrespective
of the value of $J'$, as long as $J'>0$.
Increasing the electron concentration for $J'>0.02$, the system favors
first a C-phase, then an A-phase, and finally ferromagnetism (F).
Similar results have been found for the
$J_H=\infty$ limit~\cite{brink99:_DE_two_orbital}. Finite values for $J_H$ have almost negligible
influence on the phase diagram for densities $n<0.5$. The G-phase has not been
reported in \cite{brink99:_DE_two_orbital} since it has not been  taken into consideration.
For small super-exchange of the $t_{2g}$ spins ($J'<0.018$) the transition from G to A phase evolves
via spin canting. By increasing the electron doping, the F-phase is reached without canting. The
situation is more complex for larger values of $J'$. Fig.~\ref{tztxy_vs_y_J8S025.eps} shows the
optimal angles $\theta_z$ and $\theta_{xy}$ as a function of the electron density $n$ for $J'=0.025$
and $J_H=8$. For $n<0.039$, there is 3D AF order (G-type) with an increasing tendency towards
canting between $t_{2g}$ spins in the $xy$ plane. At $n\simeq0.04$ this tendency is strongly reduced
while at the same time $\cos(\theta_{z})$ discontinuously jumps to zero and we gradually enter
the C-phase by aligning the spins along the chains in $z$-direction. The
ferromagnetic chains are not perfectly antiferromagnetically stacked. At about $8\%$ electron
density  we observe a phase transition to the A-phase. At $n\approx 0.15$ an abrupt transition to 3D
ferromagnetic order occurs.

Besides the analysis for the canonical ensemble, computations have been
performed for the {\em grand canonical ensemble} as well.
The results are significantly different. In the grand canonical ensemble
only the $G$ and the $F$ phase remain. The white solid line in Fig.~\ref{phases3d_J8.eps}
represents the phase boundary between the two phases.
For fixed $J'$, the behavior is similar to
that of the 1D system, depicted in Fig.~\ref{mu_n_L20Jx.eps}.
Below a critical chemical potential $\mu^c_1$ the electron density is zero and the $t_{2g}$ spins
have AF order. At $\mu=\mu^c_1$, zero density and a finite density $n_1$, given by
the solid line in Fig.~\ref{phases3d_J8.eps}, coexist.
Concurrently with phase separation, AF- and FM order coexist.
Above $\mu^c_1$ the density increases monotonically.
The second transition at $\mu^c_2$ is not shown in Fig.~\ref{phases3d_J8.eps}
as it occurs close to $n=1$.
Therefore, the grand canonical UHA result does not exhibit the additional magnetic
phases (A and C), which are observed in experiment. The relevant densities
are never stabilized.

\section{MC simulations for 1D systems}                          \label{sec:results}

In this section we compare MC results, obtained for the
original double exchange (DE) model~(\ref{eq:H}),
with those for the effective spinless fermion model (\Eq{eq:Hp}), where the
additional (``three-site'') hopping term
has been neglected. We use the grand canonical Monte
Carlo method introduced in~\cite{yunoki98:_phase} with open boundary conditions (obc).

We restrict the discussion to 1D systems. There is no reason to believe that the performance of the
approximation of effective spinless fermions
is different in higher dimensions. Furthermore, the approximation
has little influence on the orbital
degrees of freedom and we restrict the analysis to the one-orbital model.
In all simulations, the super-exchange coupling of the $t_{2g}$ spins is $J'=0.02$.

Fig.~\ref{mu_n_L20Jx.eps} shows the dependence of the electron density on the chemical potential.
The system parameters are $L=20$, $\beta=50$, and $J_H$ is varied between $J_H=4$ and $J_H=10$. All
results for the two models are in almost perfect agreement. The 'largest' difference can be
observed at $\mu^c_2$ for $J_H=4$. The lines in Fig.\ \ref{mu_n_L20Jx.eps} represent the
results of the uniform hopping approximation, which are strikingly close to the MC data points.
The discontinuities  are more pronounced in UHA than in the MC data, which can partially be
attributed to the fact that the UHA results are for $L=\infty$ and $T=0$.
The treatment of finite temperatures in the UHA requires the determination of
the number of $t_{2g}$ configurations at a given $u$ and is the subject of
current investigations \cite{kol:pru:wvl:2}.

\begin{figure}[!ht]
\includegraphics[width=0.4\textwidth]{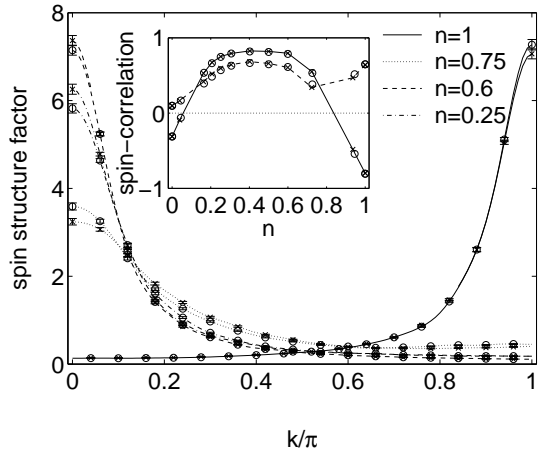}
\caption{
  Structure factor for the $t_{2g}$ spins at various electron densities with $J_H=6$,
  $\beta=50$, and $L=20$, for the spinless fermion model $H_p$ (circles),
  and for the  DE model $H$ (crosses).
  When not shown, error bars are smaller than the symbols.
  The inset shows the nearest neighbor (solid line) and next nearest neighbor
  (dashed line)  $t_{2g}$ spin correlation.
} \label{p2_dg_spin_J6.eps}
\end{figure}

\begin{figure}[!ht]
\includegraphics[width=0.4\textwidth]{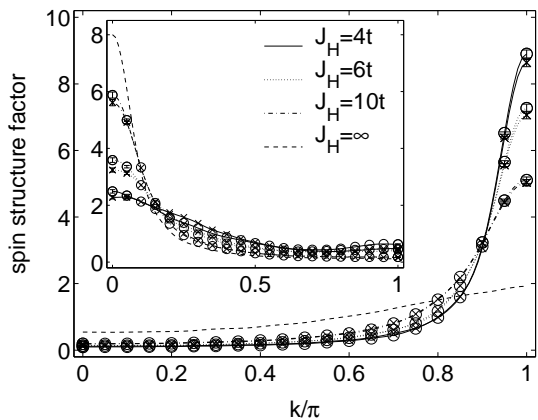}
\caption{Spin structure factor at $n=1$ (inset: $n\approx 0.75$) for
  different values of $J_H$.
  Same symbols and parameters as in Fig.~\ref{p2_dg_spin_J6.eps}.
  In the limit $J_H\to \infty$ (dashed line) the intensity of the AF peak is considerably
  reduced.
} \label{spin_structure_L20NxJx.eps}
\end{figure}

The structure factor of the $t_{2g}$ spins has been calculated for various densities $n$
in the grand canonical ensemble by adjusting the chemical potential.
The results are illustrated in Fig.~\ref{p2_dg_spin_J6.eps}.
Again the data for the two models,
Kondo- and effective spinless fermion model, are in perfect
agreement within the error bars.
Corroborating the UHA results of the previous section, the filled band ($n=1$)
has a peak at $k=\pi$, corresponding to AF order. For decreasing density
the ferromagnetic peak increases up to $n=1/2$ and then it decreases again.
The inset  shows the nearest neighbor and next nearest neighbor
spin correlation function versus density. We observe  that both models yield the same magnetic
behavior: AF order at low and high density and a ferromagnetic phase at intermediate fillings. The
pronounced peak at $k=\pi$ for $n=1$ results from (virtual) spin-flip processes,
driven by the relatively strong exchange coupling
$J_{\text{eff}}=0.103$, \Eq{Jeff}.
Contrariwise, the AF structure near $n=0$ is much less pronounced as it is
merely driven by the weak super-exchange coupling $J'=0.02$.
Fig.~\ref{spin_structure_L20NxJx.eps} shows the structure factor
for the $t_{2g}$ spins at $n=1$ for different
values of $J_H$. The AF peak decreases with increasing $J_H$ and degenerates to a broad structure
in the limit $J_H \to \infty$.
Obviously, the inclusion of the second order term to the effective spinless fermion model,
which is missing in the commonly used $J_H=\infty$ limit,
and which provides the strong exchange coupling $J_{\text{eff}}$,
is crucial for the correct description of the AF order at high electron density.
The inset shows the spin structure for $n\approx3/4$, at which the system
exhibits ferromagnetic order.
In this case, the FM correlations increase with increasing $J_H$.
\begin{figure}[!ht]
\includegraphics[width=0.3\textwidth]{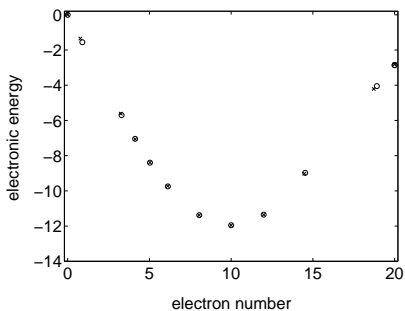}
\caption{Electronic contribution to the total energy versus electron number for $\beta=50$, $L=20$,
and $J_H=6$. Results for the DE model $\hat{H}$ (crosses)  are compared with those for the
spinless fermion model $\hat{H}_p$ (circles).} \label{dg_mf_energy.eps}
\end{figure}
As a further test for the spinless fermion model, the electronic contribution
to the total energy is shown in Fig.~\ref{dg_mf_energy.eps}.
The energy for the Kondo-lattice model (first two terms in \Eq{eq:H})  is compared with
the corresponding contributions in the effective spinless fermion
model (\Eq{eq:Hp}). For all fillings the results are in very good
agreement.

\begin{figure}[!ht]
\includegraphics[width=0.3\textwidth]{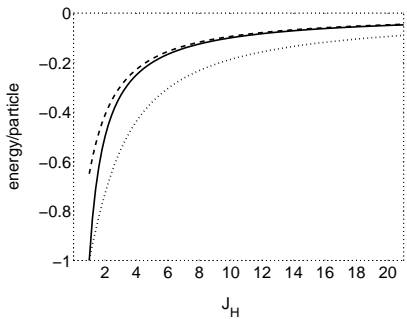}
\caption{
  Electronic energies versus Hund coupling $J_H$ for a completely filled $20$-site chain at
  $T=0$.   Kinetic energy (dotted line) and total electronic energy (dashed lines)
  for the Kondo-lattice model.
  The solid line represents the total electronic energy for the effective
  spinless fermion model.}
\label{dg_mf_energydiff.eps}
\end{figure}

Quantitatively, the largest differences are found for $n=1$.
For this density the dependence of the  energy on $J_H$ is studied in Fig.~\ref{dg_mf_energydiff.eps}.
A detailed comparison reveals
that the effective model describes the electronic energy extremely well, even
in the moderate coupling regime.
The ubiquitous $J_H=\infty$ approximation, on the other hand, yields zero electronic energy.
For the parameters underlying Fig.~\ref{dg_mf_energydiff.eps},
the $t_{2g}$ spins are antiferromagnetically  ordered and
the lower Kondo-band, or rather the single band of the spinless
fermion model, is completely filled. Nevertheless, the kinetic electronic energy is finite
due to (virtual) spin-flip processes.
It should be pointed out that the additional (three-site) hopping term in \Eq{eq:Hp}
has no influence on this result as the band is entirely filled.

\begin{figure}[!ht]
\includegraphics[width=0.3\textwidth]{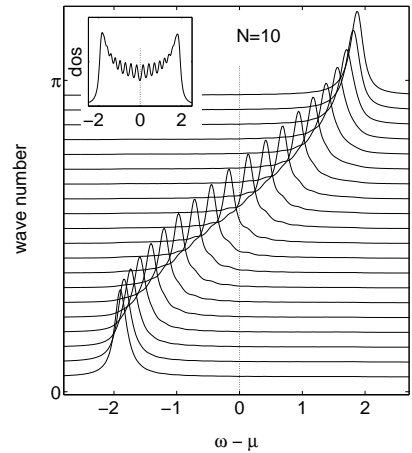}
\caption{Spectral function
obtained by a grand canonical simulation
for $n=1/2$ with $J_H=6$, $L=20$ and $\beta=50$.
An intrinsic linewidth $\gamma=0.1$ has been added.
The results for the Kondo model and the effective spinless fermion model are
indistinguishable.
}
\label{fig:spectra_J6_N10}
\end{figure}

Next we study the spectral function $A_k(\omega)$ in the grand canonical ensemble
for various mean electron densities
covering the regimes for AF and FM order, as well as phase separation. In all cases
the system geometry is a 20-site chain with open boundary conditions at inverse temperature
$\beta=50$, and exchange couplings $J'=0.02$, and $J_H=6$.
We start out with the spectra in Fig.~\ref{fig:spectra_J6_N10}
for strong FM order at a mean particle density of $n=1/2$.
According to the inset of Fig.~\ref{p2_dg_spin_J6.eps} the
spin-correlations are 0.82 and 0.67 for
nearest and next-nearest neighbors, respectively.
The spectral function, depicted in Fig.\ \ref{fig:spectra_J6_N10}, resembles closely
that of a tight-binding model, valid for
perfect FM order.
The band-width is slightly reduced,
and for $k$-values close to the Fermi momentum, $A_k(\omega)$ exhibits some minor
shoulders. The width (HWHM) of the peaks agrees with $\gamma$, the value by which the finite-size
delta peaks have been broadened. The inset displays the density of states (dos),
which agrees with the tight-binding
density of states for open boundary conditions.
\begin{figure}
\includegraphics[width=0.3\textwidth]{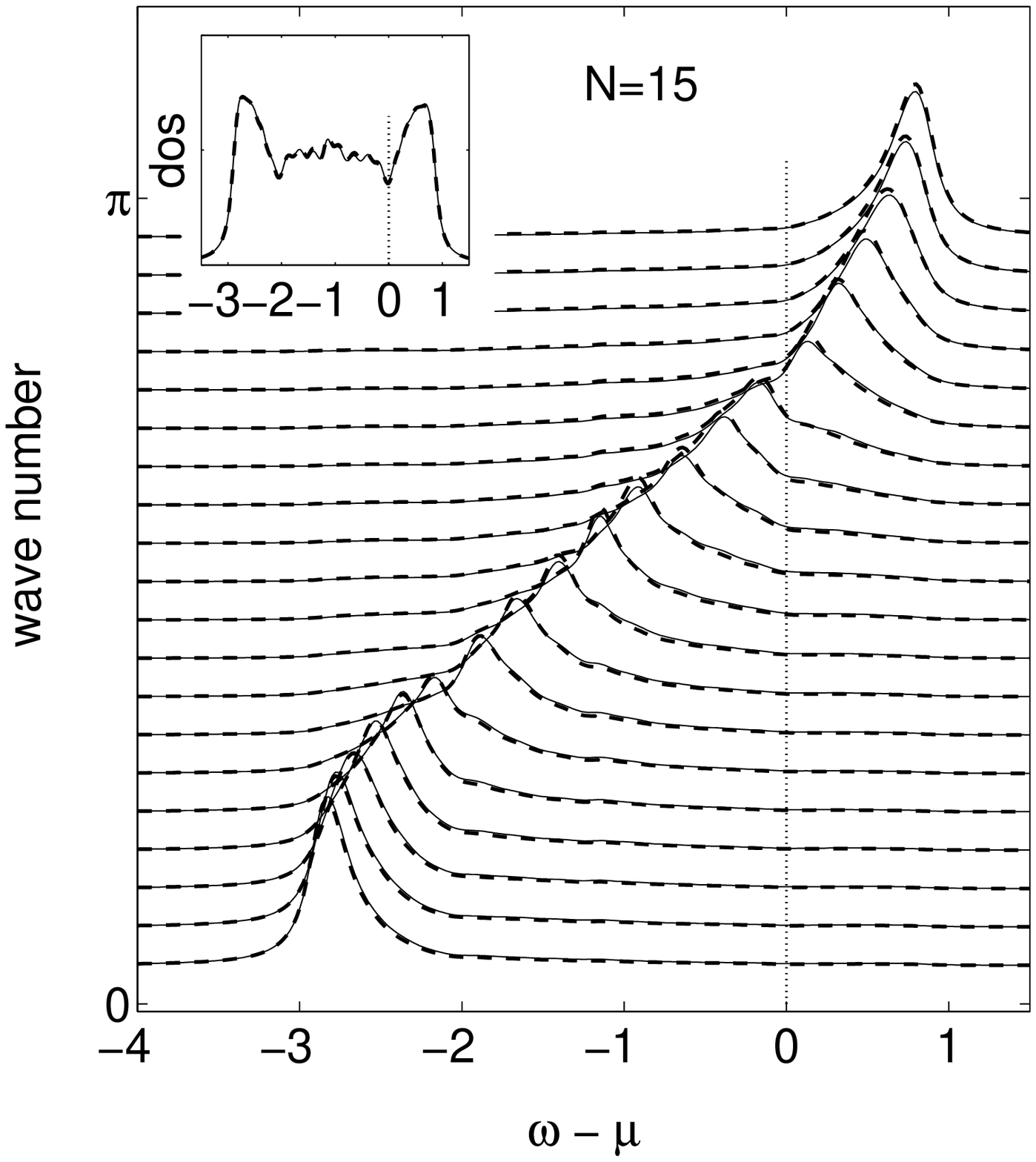}\\
\includegraphics[width=0.3\textwidth]{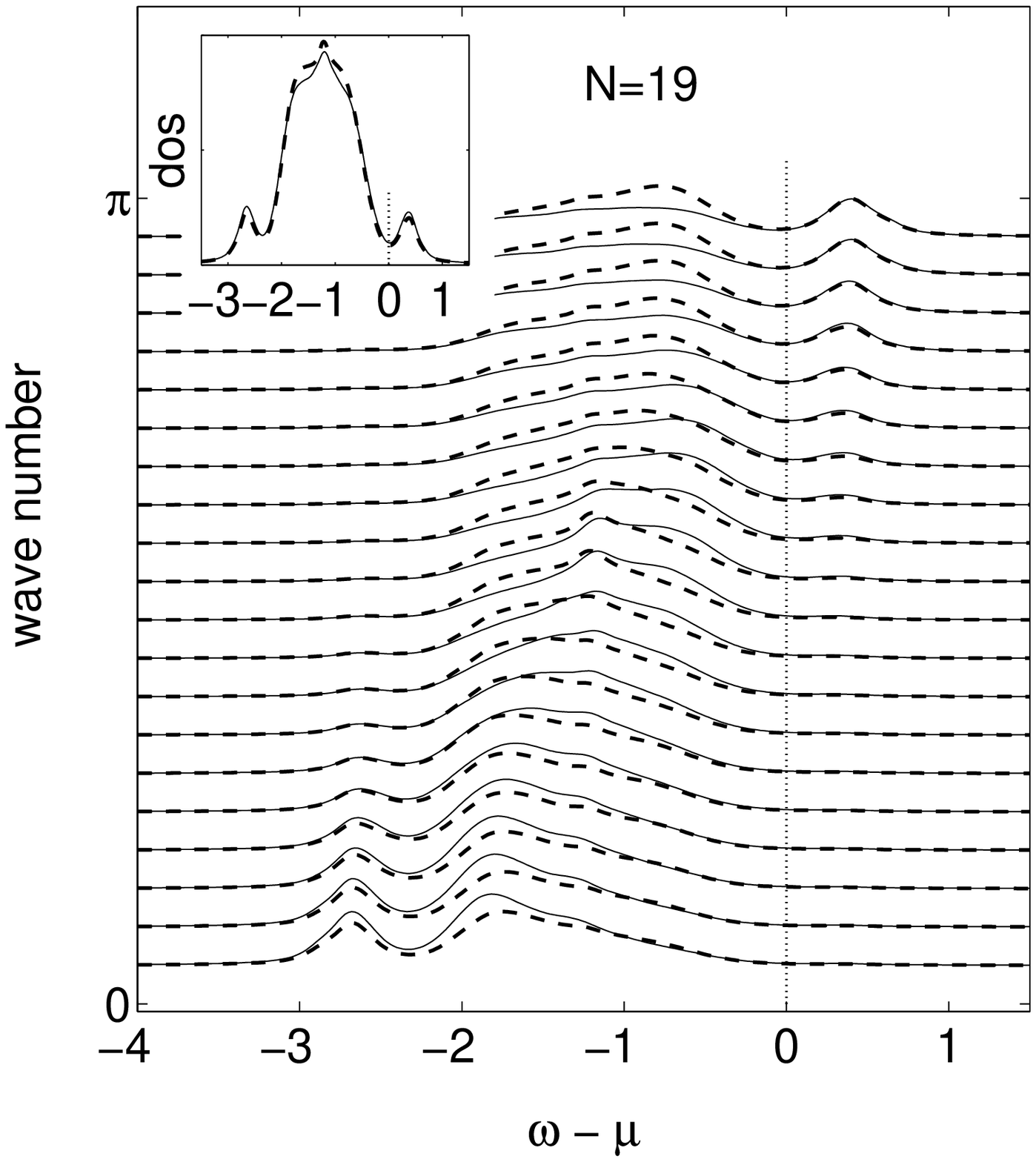}\\
\includegraphics[width=0.3\textwidth]{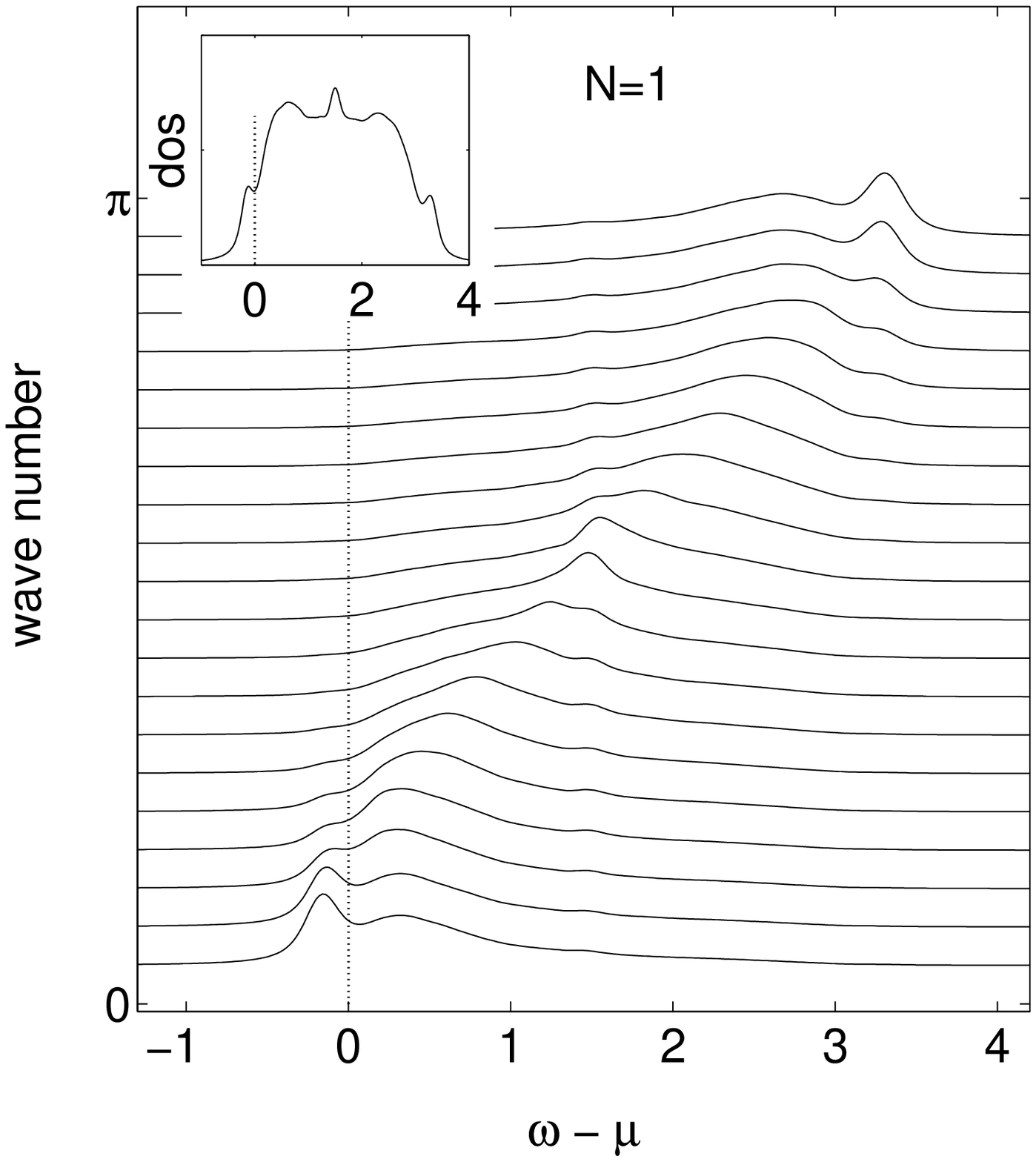}
\caption{Same as Fig.\ \ref{fig:spectra_J6_N10}.
The indicated particle numbers correspond to $n = 0.75$ $n=0.95$, and $n=0.05$.
Results for the Kondo model are represented by solid lines and those for the effective spinless
fermion model by dashed lines.
} \label{fig:spectra}
\end{figure}
Next we increase the mean density to $n=0.75$, corresponding to a chemical potential close
to $\mu^c_2$.
The spin order is still predominantly ferromagnetic.
The results in Fig.~\ref{fig:spectra} show that both models yield very similar results, namely a
tight-binding type of quasi-particle band.
The spectral peaks are, however, significantly broader than the mock width $\gamma$
and upon approaching the Fermi momentum the width increases asymmetrically towards the
Fermi level.
The origin of the broad structures are the random
deviations of the $t_{2g}$ spins from perfect FM order.
The resulting density of states has piled up spectral weight in the center and reveals a precursor
of a pseudo-gap at the Fermi energy.
Interestingly, the dos is almost center-symmetric and a mirror
image of the 'pseudo-gap' occurs also at the lower band edge.

The next panel depicts results for $n=0.95$,  correspond to a chemical potential slightly above
$\mu_{c2}$, where the spin order is predominantly antiferromagnetic. Now the pseudo-gap,
as discussed by Moreo et al. \cite{moreo99},
is clearly visible. Additionally, we observe a 'mirror pseudo-gap' at the lower band edge.
There is still good qualitative agreement between the
results of the two models. Quantitatively, however, there are deviations in the structures below
the pseudo-gap. The density of states is still remarkably well described by the spinless fermion
model.

In the opposite limit of low carrier concentration ($n=0.05$), the chemical potential is
close to $\mu_{c1}$, where we find similar features.
Again, together with the coexistence of FM and AFM order
a pseudo-gap shows up at the chemical potential as well as at the upper band
edge.
The pseudo-gap is less pronounced in this case, where the antiferromagnetic exchange coupling
$J'$ is much smaller than $J_{\text{eff}}$ at $\mu^c_2$.

In both models a considerable amount of spectral weight is transferred from the
band edges to the center.  Interestingly, the density of states is almost
center-symmetric. The pseudo-gap is present in the spectral function irrespective of the
wave vector $k$.
Moreo et al.\ argued that the pseudo-gap is formed due to the presence of mixed phases with
irregular formations of FM domains.
In contrast to this interpretation, we find the pseudo-gap also in the perfect AF regime with a single
electron (hole).

Generally we observe, in agreement with ARPES experiments \cite{dessau99}, that the width of peaks
increases towards the Fermi energy and the spectral intensity decreases since spectral weight is
transferred to the unoccupied part of the spectrum, which is not visible in ARPES. Furthermore,
the peaks are generally much broader than the experimental resolution.

\section{Conclusions}                                      \label{sec:conclusions}

We have developed an effective spinless fermion model for the strong coupling
multi-orbital Kondo-lattice model.
The effective model has a reduced Hilbert space and is particularly suitable
for MC simulations. The numerical complexity is the same as that of the $J_H=\infty$ model.
The reduced Hilbert space allows one to study
higher spatial dimensions and/or  additional degrees of freedom, such as phonons.

Based on the evaluation of various observables the effective spinless fermion model
performs strikingly well, even for moderate Hund coupling.

It appears that virtual spin-flip processes included in our approach, which are missing in the
$J_H=\infty$ model, are crucial for the antiferromagnetic phase close to half-filling ($n=1$),
where they provide a strong effective exchange coupling
$J_{\text{eff}}=J' + \frac{1}{2J_H}$.

Two phase transitions from AF- to FM-order and vice versa are observed,
accompanied by phase separation.
Analytic expressions for the chemical potential at which phase separation
occurs in a 1D chain have been derived in uniform hopping approximation (UHA).
It has been shown, however, that they are in extremely good agreement with
approximation-free MC results.

The UHA phase diagram of the 3D spinless fermion model has been determined.
In canonical ensembles,
the magnetic phase diagram is in qualitative agreement with that obtained in previous studies
for the $J_H=\infty$ limit.
Experimentally observed phases, like G-, A-, C-, and F-order are found.
On the other hand, grand canonical ensemble calculations show that only 3D AF and 3D FM
order prevail. The transition between the two phases is accompanied by phase separation.
Densities, required for other phases, are not stable in UHA.

The spectral functions show a remarkable center-symmetry. In the AF phase,
at low and high electron density, a pseudo-gap structure is observed at the chemical potential and
a mirror image at the opposite edge of the spectrum.

In passing, it should be noted that
nearest-neighbor Coulomb repulsion of the $e_g$-electrons
in a two-orbital model can be detrimental for
phase separation. Detailed results will be discussed elsewhere~\cite{kol:pru:wvl:2}.

\section{Acknowledgments}                                      \label{sec:Acknowledgments}
This work was partially supported by the Austrian Science Fund (FWF), project
P15834-PHY.


\begin{thebibliography}{20}
\expandafter\ifx\csname natexlab\endcsname\relax\def\natexlab#1{#1}\fi
\expandafter\ifx\csname bibnamefont\endcsname\relax
  \def\bibnamefont#1{#1}\fi
\expandafter\ifx\csname bibfnamefont\endcsname\relax
  \def\bibfnamefont#1{#1}\fi
\expandafter\ifx\csname citenamefont\endcsname\relax
  \def\citenamefont#1{#1}\fi
\expandafter\ifx\csname url\endcsname\relax
  \def\url#1{\texttt{#1}}\fi
\expandafter\ifx\csname urlprefix\endcsname\relax\def\urlprefix{URL }\fi
\providecommand{\bibinfo}[2]{#2}
\providecommand{\eprint}[2][]{\url{#2}}

\bibitem[{\citenamefont{Dagotto et~al.}(2001)\citenamefont{Dagotto, Hotta, and
  Moreo}}]{dagotto01:review}
\bibinfo{author}{\bibfnamefont{E.}~\bibnamefont{Dagotto}},
  \bibinfo{author}{\bibfnamefont{T.}~\bibnamefont{Hotta}}, \bibnamefont{and}
  \bibinfo{author}{\bibfnamefont{A.}~\bibnamefont{Moreo}},
  \bibinfo{journal}{Phys. Reports} \textbf{\bibinfo{volume}{344}},
  \bibinfo{pages}{1} (\bibinfo{year}{2001}).

\bibitem[{\citenamefont{Ole\'{s} et~al.}(2000)\citenamefont{Ole\'{s}, Cuoco,
  and Perkins}}]{oles00:_magnet_orbit_order_cuprat_mangan}
\bibinfo{author}{\bibfnamefont{A.~M.} \bibnamefont{Ole\'{s}}},
  \bibinfo{author}{\bibfnamefont{M.}~\bibnamefont{Cuoco}}, \bibnamefont{and}
  \bibinfo{author}{\bibfnamefont{N.~B.} \bibnamefont{Perkins}}, in
  \emph{\bibinfo{booktitle}{AIP Conference Proceedings}}
  (\bibinfo{year}{2000}), vol. \bibinfo{volume}{527}, pp.
  \bibinfo{pages}{226--380}.

\bibitem[{\citenamefont{Kaplan and Mahanti}(1998)}]{proceedings98}
\bibinfo{author}{\bibfnamefont{T.}~\bibnamefont{Kaplan}} \bibnamefont{and}
  \bibinfo{author}{\bibfnamefont{S.}~\bibnamefont{Mahanti}},
  \emph{\bibinfo{title}{Physics of Manganites}} (\bibinfo{publisher}{Kluwer
  Academic/ Plenum Publishers}, \bibinfo{address}{New York, Boston, Dordrecht,
  London, Moscow}, \bibinfo{year}{1998}), \bibinfo{edition}{1st} ed.

\bibitem[{\citenamefont{Horsch et~al.}(1999)\citenamefont{Horsch, Jaklic, and
  Mack}}]{horsch99}
\bibinfo{author}{\bibfnamefont{P.}~\bibnamefont{Horsch}},
  \bibinfo{author}{\bibfnamefont{J.}~\bibnamefont{Jaklic}}, \bibnamefont{and}
  \bibinfo{author}{\bibfnamefont{F.}~\bibnamefont{Mack}},
  \bibinfo{journal}{Phys. Rev. B} \textbf{\bibinfo{volume}{59}},
  \bibinfo{pages}{R14149} (\bibinfo{year}{1999}).

\bibitem[{\citenamefont{Bala et~al.}(2002)\citenamefont{Bala, Oles, and
  Horsch}}]{bala02}
\bibinfo{author}{\bibfnamefont{J.}~\bibnamefont{Bala}},
  \bibinfo{author}{\bibfnamefont{A.~M.} \bibnamefont{Oles}}, \bibnamefont{and}
  \bibinfo{author}{\bibfnamefont{P.}~\bibnamefont{Horsch}},
  \bibinfo{journal}{Phys. Rev. B} \textbf{\bibinfo{volume}{65}},
  \bibinfo{pages}{134420/1} (\bibinfo{year}{2002}).

\bibitem[{\citenamefont{Zener}(1951)}]{zener51}
\bibinfo{author}{\bibfnamefont{C.}~\bibnamefont{Zener}},
  \bibinfo{journal}{Phys. Rev.} \textbf{\bibinfo{volume}{82}},
  \bibinfo{pages}{403} (\bibinfo{year}{1951}).

\bibitem[{\citenamefont{Dagotto et~al.}(1998)\citenamefont{Dagotto, Yunoki,
  Malvezzi, Moreo, Hu, Capponi, Poilblanc, and
  Furukawa}}]{dagotto98:_ferrom_kondo_model_mangan}
\bibinfo{author}{\bibfnamefont{E.}~\bibnamefont{Dagotto}},
  \bibinfo{author}{\bibfnamefont{S.}~\bibnamefont{Yunoki}},
  \bibinfo{author}{\bibfnamefont{A.~L.} \bibnamefont{Malvezzi}},
  \bibinfo{author}{\bibfnamefont{A.}~\bibnamefont{Moreo}},
  \bibinfo{author}{\bibfnamefont{J.}~\bibnamefont{Hu}},
  \bibinfo{author}{\bibfnamefont{S.}~\bibnamefont{Capponi}},
  \bibinfo{author}{\bibfnamefont{D.}~\bibnamefont{Poilblanc}},
  \bibnamefont{and} \bibinfo{author}{\bibfnamefont{N.}~\bibnamefont{Furukawa}},
  \bibinfo{journal}{Phys. Rev. B} \textbf{\bibinfo{volume}{58}},
  \bibinfo{pages}{6414} (\bibinfo{year}{1998}).

\bibitem[{\citenamefont{Furukawa}(1998)}]{furukawa98}
\bibinfo{author}{\bibfnamefont{N.}~\bibnamefont{Furukawa}},
  \emph{\bibinfo{title}{in: Physics of manganites}} (\bibinfo{publisher}{Kluwer
  Academic Publisher}, \bibinfo{address}{New York}, \bibinfo{year}{1998}),
  \bibinfo{edition}{1st} ed.

\bibitem[{\citenamefont{Yunoki and
  Moreo}(1998)}]{yunoki98:_static_dynam_proper_ferrom_kondo}
\bibinfo{author}{\bibfnamefont{S.}~\bibnamefont{Yunoki}} \bibnamefont{and}
  \bibinfo{author}{\bibfnamefont{A.}~\bibnamefont{Moreo}},
  \bibinfo{journal}{Phys. Rev. B} \textbf{\bibinfo{volume}{58}},
  \bibinfo{pages}{6403} (\bibinfo{year}{1998}).

\bibitem[{\citenamefont{Yunoki et~al.}(1998{\natexlab{a}})\citenamefont{Yunoki,
  Hu, Malvezzi, Moreo, Furukawa, and Dagotto}}]{yunoki98:_phase}
\bibinfo{author}{\bibfnamefont{S.}~\bibnamefont{Yunoki}},
  \bibinfo{author}{\bibfnamefont{J.}~\bibnamefont{Hu}},
  \bibinfo{author}{\bibfnamefont{A.~L.} \bibnamefont{Malvezzi}},
  \bibinfo{author}{\bibfnamefont{A.}~\bibnamefont{Moreo}},
  \bibinfo{author}{\bibfnamefont{N.}~\bibnamefont{Furukawa}}, \bibnamefont{and}
  \bibinfo{author}{\bibfnamefont{E.}~\bibnamefont{Dagotto}},
  \bibinfo{journal}{Phys. Rev. Lett.} \textbf{\bibinfo{volume}{80}},
  \bibinfo{pages}{845} (\bibinfo{year}{1998}{\natexlab{a}}).

\bibitem[{\citenamefont{Hotta et~al.}(2000)\citenamefont{Hotta, Malvezzi, , and
  Dagotto}}]{hotta00:coo_ps_nn_coulomb}
\bibinfo{author}{\bibfnamefont{T.}~\bibnamefont{Hotta}},
  \bibinfo{author}{\bibfnamefont{A.~L.} \bibnamefont{Malvezzi}}, ,
  \bibnamefont{and} \bibinfo{author}{\bibfnamefont{E.}~\bibnamefont{Dagotto}},
  \bibinfo{journal}{Phys. Rev. B} \textbf{\bibinfo{volume}{62}},
  \bibinfo{pages}{9432} (\bibinfo{year}{2000}).

\bibitem[{\citenamefont{Yunoki et~al.}(1998{\natexlab{b}})\citenamefont{Yunoki,
  Moreo, and Dagotto}}]{yunoki98:_phase_separ_induc_orbit_degrees}
\bibinfo{author}{\bibfnamefont{S.}~\bibnamefont{Yunoki}},
  \bibinfo{author}{\bibfnamefont{A.}~\bibnamefont{Moreo}}, \bibnamefont{and}
  \bibinfo{author}{\bibfnamefont{E.}~\bibnamefont{Dagotto}},
  \bibinfo{journal}{Phys. Rev. Lett.} \textbf{\bibinfo{volume}{81}},
  \bibinfo{pages}{5612} (\bibinfo{year}{1998}{\natexlab{b}}).

\bibitem[{\citenamefont{van~den Brink and
  Khomskii}(1999)}]{brink99:_DE_two_orbital}
\bibinfo{author}{\bibfnamefont{J.}~\bibnamefont{van~den Brink}}
  \bibnamefont{and} \bibinfo{author}{\bibfnamefont{D.}~\bibnamefont{Khomskii}},
  \bibinfo{journal}{Phys. Rev. Lett} \textbf{\bibinfo{volume}{82}},
  \bibinfo{pages}{1016} (\bibinfo{year}{1999}).

\bibitem[{\citenamefont{von~der Linden and Nolting}(1982)}]{wvdl82}
\bibinfo{author}{\bibfnamefont{W.}~\bibnamefont{von~der Linden}}
  \bibnamefont{and} \bibinfo{author}{\bibfnamefont{W.}~\bibnamefont{Nolting}},
  \bibinfo{journal}{Z. Phys. B} \textbf{\bibinfo{volume}{48}},
  \bibinfo{pages}{191} (\bibinfo{year}{1982}).

\bibitem[{\citenamefont{Auerbach}(1994)}]{Auerbach:book}
\bibinfo{author}{\bibfnamefont{A.}~\bibnamefont{Auerbach}},
  \emph{\bibinfo{title}{Interacting Electrons and Quantum Magnetism}}
  (\bibinfo{publisher}{Springer-Verlag}, \bibinfo{address}{New York, Berlin,
  Heidelberg}, \bibinfo{year}{1994}), \bibinfo{edition}{1st} ed.

\bibitem[{\citenamefont{Yarlagadda and Ting}(2001)}]{yarlagadda01:mf_COSO}
\bibinfo{author}{\bibfnamefont{S.}~\bibnamefont{Yarlagadda}} \bibnamefont{and}
  \bibinfo{author}{\bibfnamefont{C.~S.} \bibnamefont{Ting}},
  \bibinfo{journal}{Int. J. Mod. Phys. B} \textbf{\bibinfo{volume}{15}},
  \bibinfo{pages}{2719} (\bibinfo{year}{2001}).

\bibitem[{\citenamefont{Shen and Wang}(2000)}]{SQShen00:pp_mf}
\bibinfo{author}{\bibfnamefont{S.-Q.} \bibnamefont{Shen}} \bibnamefont{and}
  \bibinfo{author}{\bibfnamefont{Z.~D.} \bibnamefont{Wang}},
  \bibinfo{journal}{Phys. Rev. B} \textbf{\bibinfo{volume}{61}},
  \bibinfo{pages}{9532} (\bibinfo{year}{2000}).

\bibitem[{\citenamefont{Koller et~al.}(2002)\citenamefont{Koller, Pr\"ull,
  Evertz, and von~der Linden}}]{kol:pru:wvl:2}
\bibinfo{author}{\bibfnamefont{W.}~\bibnamefont{Koller}},
  \bibinfo{author}{\bibfnamefont{A.}~\bibnamefont{Pr\"ull}},
  \bibinfo{author}{\bibfnamefont{H.~G.} \bibnamefont{Evertz}},
  \bibnamefont{and} \bibinfo{author}{\bibfnamefont{W.}~\bibnamefont{von~der
  Linden}} (\bibinfo{year}{2002}), \bibinfo{note}{in preparation}.

\bibitem[{\citenamefont{Moreo et~al.}(1999)\citenamefont{Moreo, Yunoki, and
  Dagotto}}]{moreo99}
\bibinfo{author}{\bibfnamefont{A.}~\bibnamefont{Moreo}},
  \bibinfo{author}{\bibfnamefont{S.}~\bibnamefont{Yunoki}}, \bibnamefont{and}
  \bibinfo{author}{\bibfnamefont{E.}~\bibnamefont{Dagotto}},
  \bibinfo{journal}{Phys. Rev. Lett.} \textbf{\bibinfo{volume}{83}},
  \bibinfo{pages}{2773} (\bibinfo{year}{1999}).

\bibitem[{\citenamefont{Dessau et~al.}(1998)\citenamefont{Dessau, Saitoh, Park,
  Shen, Villella, Hamada, Moritomo, and Tokura}}]{dessau99}
\bibinfo{author}{\bibfnamefont{D.~S.} \bibnamefont{Dessau}},
  \bibinfo{author}{\bibfnamefont{T.}~\bibnamefont{Saitoh}},
  \bibinfo{author}{\bibfnamefont{C.~H.} \bibnamefont{Park}},
  \bibinfo{author}{\bibfnamefont{Z.~X.} \bibnamefont{Shen}},
  \bibinfo{author}{\bibfnamefont{P.}~\bibnamefont{Villella}},
  \bibinfo{author}{\bibfnamefont{N.}~\bibnamefont{Hamada}},
  \bibinfo{author}{\bibfnamefont{Y.}~\bibnamefont{Moritomo}}, \bibnamefont{and}
  \bibinfo{author}{\bibfnamefont{Y.}~\bibnamefont{Tokura}},
  \bibinfo{journal}{Phys. Rev. Lett.} \textbf{\bibinfo{volume}{81}},
  \bibinfo{pages}{192} (\bibinfo{year}{1998}).

\end{thebibliography}

\end{document}